\documentclass[a4paper,12pt,onecolumn,final,notitlepage,oneside]{article}
\usepackage[english]{babel}
\usepackage[cp1251]{inputenc}
\usepackage[T2A]{fontenc}
\usepackage{amsmath}%
\usepackage{amsfonts}%
\usepackage{amssymb}%
\begin{document}
\title{The General Form-Invariance Principle}
\author{V. V. Voytik\thanks {voytik1@yandex.ru}}
\maketitle
\begin{center}
Bashkir State Pedagogical University,

ul. Oktyabr’skoi revoliutsii, 3a, Ufa, 450077, Russia

Received December 15, 2010; in final form March 7, 2011;

Published in the journal Gravitation and Cosmology, 2011,

 Vol. 17, No. 3, pp. 218–223
\end{center}
\begin{abstract}
We postulate the applicability of the general form-invariance principle
  in special relativity. It is shown that this principle holds in classical
  mechanics. Some examples of transformations between the reference frames
  which satisfy this principle are considered. A new transformation is
  proposed for a transition from a uniformly rotating reference frame
  to a reference frame whose origin is shifted from the rotation axis.
  Another possible formulation of this principle is given for the case of
  stationary rigid reference frames which differ from one another
  only by the position of the origin.
\end{abstract}



\section{Introduction}

The content of special relativity (SR) is to a greatest extent determined
  by the relativity principle. Therefore its different formulations are
  very important. The Poincar\'e--Einstein special relativity principle,
  corresponding exclusively to inertial reference frames, underwent a
  certain evolution from its formulation in 1904 to the time when it became
  a general principle. Historically, the general relativity principle is
  closely related to the equivalence principle and the general covariance
  condition. At first there was no clear distinction between these notions.
  The first clear formulation of the general relativity principle as an
  independent statement that {\it the laws of nature should be formulated
  in such a way that they hold with respect to arbitrarily moving coordinate
  systems\/} has appeared only in 1915 in the paper ``Relativity Theory''.
  However, then and later the author saw the value of such an
  assumption in the equivalence principle (EP), i.e., in the fact that it
  enables one to replace a homogeneous field of gravity with a uniformly
  accelerated reference frame. The general covariance condition became a
  mathematical expression of this requirement.
  
  The final formulation of the general relativity principle appeared in the
  work of 1916 ``Foundations of General Relativity'' \cite {1} and did not
  substantially change afterwards. According to this principle, the
  equations that express the laws of nature must be generally covariant
  with respect to all continuous coordinate transformations. {\it The
  general laws of nature should be expressed through equations valid in all
  coordinate systems, i.e. these equations should be covariant with respect
  to anty substitutions (generally covariant)\/}.

  It is necessary to mention that there is at present \cite {2} another,
  more general formulation of the special relativity principle, which is
  slightly weaker than Einstein's general relativity principle. It has
  received the name of the generalized relativity principle,
  which declares that ``{\it whatever physical reference frame we choose
  (inertial or non-inertial), one can always point out an infinite set of
  other reference frames such that all physical phenomena proceed in the
  same way with the initial reference frame; so that we do not have and
  cannot have any experimental opportunity to find out in which particular
  reference frame from this infinite set we are\/}; for any non-inertial
  reference frames, there is an infinite set of other non-inertial frames in
  which all laws of physics are absolutely the same as in the initial system.

  The main purpose of the present paper is as follows.

  Above all, it is necessary to pay attention to the circumstance that the
  general relativity principle as it was initially formulated in 1915 has a
  certain value irrespective of the opportunity of replacing the field of
  inertia forces in a non-inertial reference frame with a gravitational
  field. The principle is applicable to both classical and relativistic
  phenomena and is one of the fundamentals of SR in non-inertial reference
  frames. Further on, in Section \ref {cv} we will give another, more rigorous
  formulation, and in Section \ref {vb} we will show that this principle originates
  from classical mechanics.

  Moreover, if this principle holds, it is necessary to give examples of
  transformations between non-inertial reference frames that satisfy this
  principle. Such examples are considered in Section \ref {bn}. In particular,
  we suggest a transformation of displacement from a uniformly rotating
  reference frame to a finite distance, which is in a certain sense similar
  to the Lorentz transformations.

  Finally, in Section \ref {nm}, another possible formulation of the general
  form-invariance principle will be given, specially for the case of
  stationary rigid reference frames which differ from each other only
  by the position of the origin.

\section{The general form-invariance principle for rigid reference frames}
\label{cv}

 The general relativity principle actually states that if one takes into
  account in some way the non-inertial motion of two different reference
  frames, then the mathematical form of the physical laws will remain
  invariable at transition from one frame to the other. The question is, how
  to do it in practice. The arbitrariness of the space-time metric in a
  given reference frame can be restricted by taking into account that it
  depends only on the internal kinematic characteristics of the reference
  body. The reference body (and hence the reference frame) is entirely
  characterized by only two vectors: the proper acceleration $\mathbf{W}$
  and the proper angular velocity $\mathbf{\Omega}$, and these parameters do
  not include the coordinates of the reference body, or its velocity, or any
  other {\it a priori\/} possible quantities. All distinctions between
  reference frames are connected exclusively with differences in these
  characteristics. Thus there is no way to distinguish one non-inertial
  reference frame from another if their origins have the same kinematic
  characteristics. This implies the non-existence of a unique, absolute
  reference frame, the only one with respect to which hold the laws of
  physics.

It can be therefore concluded that if one chooses another reference frame,
  different from the initial one by the most general transformation related
  to a translational motion along with rotation, the dependence of the
  space-time metric on the Cartesian coordinates will remain unchanged. It
  means that the physical laws, as the reference frame is changed, can only
  change due to changed vector characteristics of the reference frame,
  $\mathbf{W}$  and $\mathbf{\Omega}$, and the absolute values
  of the coordinates and velocity of the reference body are insignificant:
  \begin{equation}
	g_{ik}(\mathbf{W}(T),\mathbf{\Omega}(T),\mathbf{R})=forminvariant      \label{A}
\end{equation}
  We will call this statement the {\it general form-invariance principle\/}.
   
 The condition that the space is Euclidean in an accelerated reference
  frame and the condition that the space-time is not curved uniquely fix
  the form of the metric in an arbitrary reference frame \cite {3}, \cite {4}
  (here and henceforth $c=1$):
  \begin{equation}
g_{00} =(1+\mathbf{WR})^2-(\mathbf{\Omega \times R})^{2}              \label{B}
\end{equation}
\begin{equation}
 g_{0\alpha } =-e_{\alpha \beta \gamma }\Omega_{\beta }R_{\gamma }  \label{C}
\end{equation}
\begin{equation}
 g_{\alpha \beta } =-\delta_{\alpha \beta }     \label{D}
\end{equation}
Mathematically, the general form-invariance principle means that in
  an arbitrary reference frame there exists such a transformation
  $X^i=X^i(x^j)$, or, in the three-dimensional form,
\begin{equation}
                   \mathbf{R}=\mathbf{R}(\mathbf{r},t), T=T(\mathbf{r},t)    \label{E}
               \end{equation}    
that its substitution into the metric tensor transformation law does not
  change the form of the metric:
    \begin{equation}
\frac{\partial X^{j} }{\partial x^{i} } \frac{\partial X^{l} }{\partial x^{k} } g_{jl} \label{F} (\mathbf{W,\Omega,R})=g_{ik}(\mathbf{w,\mathbf{\boldsymbol{\omega}},r})
\end{equation}
$$ds^{2} =\left\{(1+\mathbf{WR})^{2} -(\mathbf{\Omega \times R})^{2}\right\}dT^2 -2(\mathbf{\Omega \times R})d\mathbf{R}dT-d\mathbf{R}^{2} =$$
\begin{equation}
=\left\{(1+\mathbf{wr})^{2} -(\mathbf{\boldsymbol{\omega} \times r})^2\right\}dt^{2}-2(\mathbf{\boldsymbol{\omega} \times r})d\mathbf{r}dt-d\mathbf{r}^{2}  \label{G}
 \end{equation}   
 Such a transformation will be a change in the reference frame. The
  form-invariance of the metric guarantees that the physical laws will
  remain invariable as the reference frame is changed. These transformations
  evidently form a group.

  Besides translational motion and rotations, a reference frame changes at
  displacements by a fixed distance. If the initial reference frame is rigid
  and non-stationary (i.e., its characteristics are time-dependent,
  $\mathbf{W}=\mathbf{W}(T)$, $\mathbf{\Omega}=\mathbf{\Omega}(T)$), then
  the reference frame whose origin is shifted will not be rigid, and its
  metric will not have the form (\ref{B})-(\ref{D}). In this case, the question about
  the metric corresponding to such a reference frame requires an additional
  study. However, if the initial reference frame is rigid and stationary,
  the one with a shifted origin will also be rigid and stationary. For such
  a frame, (\ref{A})-(\ref{G}) will be entirely applicable as well as
  all conclusions of this section.

The statement (\ref{F}),\ (\ref{G}) is necessary for any physical theory.
  This requirement is more restrictive than that of covariance at coordinate
  transformations, and it substantially restricts the class of possible
  transformations between reference frames to only those which leave
  invariable the functional dependence of the 4-dimensional metric on the
  coordinates of the instantaneously comoving inertial reference frame. The
  values of the kinematic parameters themselves can change.

Maybe this condition does not deserve the name of a ``principle'' which is
  too obliging. But the same obvious nature is also inherent to the special
  relativity principle. But, as a matter of fact, the general
  form-invariance principle is a stronger statement than the generalized
  relativity principle. Indeed, if the vector parameters do not change as
  the reference frame is changing, then, accordingly, the numerical values
  and the functional form of the metric tensor at a given point of space do
  not change. Thus any physical phenomenon in the old and new reference
  frames, at the same initial conditions, will occur in the same way, so
  that ir will be impossible to distinguish one reference frame from
  another. It is this statement that is the essence of the generalized
  relativity principle. If the above parameters are zero, the generalized
  relativity principle passes on to the special relativity principle.

  Let us stress that the general form-invariance principle is of great
  importance for SR. Indeed, there are no other reliable statements that
  would indicate the correctness of some transformation connecting two
  reference frames. Only in the case where the coordinate and time
  transformations keep the general form (\ref{G}) it can be surely asserted that
  such a transformation between reference frames is truly related to a
  physical motion or relative dispositions.
  
\section{Validity of the general form-invariance principle for classical
     mechanics}
\label{vb}

 Let us now prove that the general form-invariance principle is valid for
  classical mechanics. To do that, it is sufficient to consider, in a
  non-inertial frame with the parameters  $\mathbf{W}$ and
  $\mathbf{\Omega}$, the simplest mechanical system, a free material point.
  It is convenient to use its Lagrangian in the form \cite{5}
\begin{equation} \label{H}
L=\frac{mU^{2} }{2} +m\mathbf{\Omega(R\times U)}+\frac{m(\mathbf{\Omega \times R)}^{2} }{2} -m\mathbf{WR}
\end{equation}

  The most general transformation of the reference frame is a combination of
  motion with a displacement, a rotation and a turn of the axes. Let us
  first carry out a transformation of motion, i.e., a transition to the
  reference frame $s$ moving to a distance $\mathbf{b}\, (t)$ with the
  velocity $\mathbf{v} = d\, \mathbf{b}/dt$. Then the coordinates
  $\mathbf{R}$ and $\mathbf{r}$ and the velocities $\mathbf{U}$ and
  $\mathbf{u}$ in the systems $S$ and $s$ are related by

\begin{equation} \label{I}
\mathbf{R=r+b},\mathbf{U=u+v}
\end{equation}

Substituting (\ref{I}) into (\ref{H}) and excluding full time derivatives, we obtain
  a new Lagrangian in the form

$$L'=\frac{mu^{2} }{2} +m\mathbf{\Omega \, (r\times u)}+\frac{m}{2} (\mathbf{\Omega \times r})^{2} -$$
\begin{equation} \label{J}
-m\, \left[\mathbf{W}+\frac{d\mathbf{v}}{dt} +\mathbf{\Omega \times v+\Omega \times (\Omega \times b)}+\left. \frac{d\mathbf{(\Omega \times b)}}{dt} \right]\right.\mathbf{r}
\end{equation}
The Lagrangian (\ref{J}) has the form (\ref{H}). The characteristics of the new
  reference frame are
\begin{equation} \label{K}
\mathbf{\boldsymbol{\omega} =\Omega} , \mathbf{w=W+\dot{v}+\dot{\Omega }\times b+2\Omega \, \, \times v+\Omega \times (\Omega \times b)}.
\end{equation}

 Thus automatically, from the validity of the general form-invariance
  condition, we have obtained the transformation laws for the
  characteristics of a reference frame at its change. From (\ref{K}) it is
  evident that the choice of another reference body, irrespective of the
  state of its motion (it can even be at rest with respect to the first one)
  leads in general to a change of the reference frame since its vector
  characteristics change.

  Let us now pass on to a reference frame $s$ rotating with respect to $S$
  with an angular velocity $\boldsymbol{\nu}$ around a common center. Then
  the coordinates $\mathbf{R}$  and $\mathbf{r}$ in the frames $S$ and $s$
  are related by the equations
\begin{equation} \label{L}
R_{\alpha } =a_{\alpha \beta } (t)\, r_{\beta }
\end{equation}
The rotation matrix $a_{\alpha \beta } $ satisfies the equalities
\begin{equation} \label{M}
a_{\alpha \beta } a_{\alpha \gamma } =\delta _{\beta \gamma } ,\frac{da_{\alpha \beta } }{dt} =e_{\beta \mu \lambda } \nu _{\lambda } a_{\alpha \mu } .
\end{equation}
where $\nu _{\alpha } $ are components of the angular velocity of the
  system $s$ in the moving coordinate system. Using a specific expression
  for the rotation matrix, e.g., in terms of the Euler angles, it is easy
  to verify the following equalities for rotation matrices, which, due to
  their form may be called the ``annihilation equalities'':
\begin{equation} \label{N}
e_{\alpha \mu \lambda } a_{\mu \beta } a_{\lambda \gamma } =e_{\mu \beta \gamma } a_{\alpha \mu } ,e_{\alpha \mu \lambda } a_{\beta \mu } a_{\gamma \lambda } =e_{\mu \beta \gamma } a_{\mu \alpha }
\end{equation}
Differentiating (\ref{L}), we obtain that the velocity $\mathbf{U}$ is equal to
\begin{equation} \label{hut}
U_{\alpha } =a_{\alpha \beta } u_{\beta } +e_{\beta \mu \lambda } \nu _{\lambda } a_{\alpha \mu } r_{\beta }
\end{equation}
Substituting (\ref{L}) and (\ref{hut}) into (\ref{H}) and using the relations (\ref{M}) and (\ref{N}) ,
  we obtain a new Lagrangian of the same form as (\ref{H}) where the
  characteristics of the new reference frame are
\begin{equation} \label{O}
w_{\alpha } =a_{\beta \alpha } W_{\beta } , \omega _{\alpha } =a_{\beta \alpha } \Omega _{\beta } +\nu _{\alpha } .
\end{equation}
Thus the validity of the general form-invariance principle for classical
  mechanics has been proved.

\section{Examples of transformations satisfying the general
    form-invariance principle}
\label{bn}

As an example of using this principle, let us find the transformation of
  displacement by a vector $\mathbf{b}$ in an accelerated reference frame
  ($\mathbf{\Omega} =0$). Let us assume that it is a usual substitution of
  the form $\mathbf{R=r+b}$. Inserting it in (\ref{G}) and making obvious
  transformations, one can reduce the interval to the form
\begin{equation} \label{P}
ds^{2} =\left(1+\frac{\mathbf{W}}{1+\mathbf{Wb}}\mathbf{r}\right)^{2} d(T+\mathbf{b}\int_0^T\mathbf{W}dT)^{2} -d\mathbf{r}^{2}
\end{equation}
The metric found from here is a special case of (\ref{B})-(\ref{D}), hence the
  assumption we made is true. In passing, we have obtained here the
  transformation of time
\begin{equation}
t=T+\mathbf{b}\int_0^T\mathbf{W}dT 
\end{equation}
From (\ref{P}) it is evident that a displacement by a vector $\mathbf{b}$
  within an accelerated reference frame changes the proper acceleration
  of the system to
\begin{equation} \label{RE}
\mathbf{w}=\frac{\mathbf{W}}{1+\mathbf{Wb}}
\end{equation}
This formula has been known for a long time \cite {6}.

  Another example of a coordinate and time transformation (\ref{E}) is the
  Lorentz-Nelson transformation of motion (it is the so-called generalized
  Lorentz transformation) which transforms an inertial reference frame
  $S(T,\mathbf{R})$ into an accelerating and rotating one, $s(t,\mathbf{r})$:
  \begin{equation}  \label{S}
 T=\frac{\mathbf{vr}}{\sqrt{1-v^{2} } } +\int _{0}^{t}\frac{dt}{\sqrt{1-v^{2}}} 
\end{equation}

\begin{equation}  \label{T}
 \mathbf{R}=\mathbf{r}+\frac{1-\sqrt{1-v^{2} } }{v^{2} \sqrt{1-v^{2} } } \mathbf{(vr)v}+\int _{0}^{t}\frac{\mathbf{v}dt}{\sqrt{1-v^{2} } }   
\end{equation}
That is, the interval $ds^2 = dT^2-d\mathbf{R}^2$ reduces to the form (\ref{G})
  where $\mathbf{w}$ is
\begin{equation}  \label{U}
\mathbf{w}=\frac{\dot{\mathbf{v}}}{\sqrt{1-v^{2} } } +\frac{1-\sqrt{1-v^{2} } }{v^{2} (1-v^{2} )} (\dot{\mathbf{v}}\mathbf{v)v} 
 \end{equation}
while $\boldsymbol{\omega}$ is the Thomas precession frequency \cite{7}
  (see also \cite{8}):
\begin{equation}  \label{Q}
\boldsymbol{\omega} =\boldsymbol{\omega}_{T} =\frac{1-\sqrt{1-v^{2} } }{v^{2} \sqrt{1-v^{2}}}\mathbf{v\times \dot{v}} 
\end{equation}
 where
\[\dot{\mathbf{v}}=\frac{d\mathbf{v}}{dt} \]

 Lastly, one more illustration of a transformation between reference frames
  is the suggested transformation of displacement by a vector
  $\mathbf{b}(b,0,0)$ with respect to an initial reference frame,
  uniformly rotating with an angular velocity $\Omega$ around the $Z$ axis
  in the Cartesian coordinate system  $S(T,X,Y,Z)$:
\begin{equation} \label{W}
 X=(b+x)\, \cos \, \frac{\Omega Vy}{\sqrt{1-V^{2} } } +\frac{y}{\sqrt{1-V^{2} } } \sin \, \frac{\Omega Vy}{\sqrt{1-V^{2} } }  
\end{equation}
\begin{equation}  \label{R}
 Y=\frac{y}{\sqrt{1-V^{2} } } \cos \, \frac{\Omega Vy}{\sqrt{1-V^{2} } } -(b+x)\; \sin \frac{\Omega Vy}{\sqrt{1-V^{2}}} 
\end{equation}
\begin{equation}
 Z=z     \label{Y}
\end{equation}
\begin{equation}
 T=\frac{t+Vy}{\sqrt{1-V^{2} } } \label{X}
\end{equation}
where
\begin{equation} \label{Z}
V=\Omega b
\end{equation}

If $V\ll 1$, $x \ll 1/\Omega$, $y \ll 1/\Omega$, the spatial transformation
  (\ref{W})-(\ref{Y}) passes over to the classical displacement transformation
\begin{equation} \label{qw}
X=b+x, Y=y, Z=z
\end{equation}
Equation (\ref{X}) has been found by H. Nikoli\`c \cite{9}. At the transformation (\ref{W})-(\ref{X}),
  the proper acceleration $\mathbf{w}$ of the new reference frame
  $s(t,x,y,z)$ is
\begin{equation} \label{we}
 w_{x} =-\frac{\Omega V}{1-V^{2} } , w_{y} =w_{z} =0
\end{equation}
and the new proper angular velocity $\boldsymbol{\omega}$ is
\begin{equation} \label{Tr}
 \omega _{z} =\frac{\Omega }{1-V^{2} } , \omega _{x} =\omega _{y} =0 
\end{equation}
This expression for the angular velocity is also known for a long time
  \cite{10}.
  
\section{The general form-invariance principle in its applications to
    stereometry and time}
\label{nm}

There are such physical laws (e.g., the least truncated action principle,
  the Fermat principle) whose mathematical formulation contains, instead
  of the space-time metric tensor, the metric tensor of space in a
  non-inertial reference frame. In this connection, we will present another,
  equivalent formulation of the general form-invariance principle for
  stationary rigid reference frames, distinct from each other only by the
  position of their origins in the coordinate system of one of them.

  Since the metric tensor remains form-invariant at displacements, also
  form-invariant must be the combination of these coefficients specifying
  the three-dimensional spatial metric and the ``metric'' of physical time.
  If the mathematical form of the spatial metric changed due to displacement,
  there would be a geometric way of choosing a certain absolute reference
  frame from the set of different stationary reference frames with the same
  characteristics, so that the spatial metric would be the simplest in terms
  of this absolute frame. This is certainly an incredible opportunity.
  In a similar way, the mathematical form of the element of physical time of
  a local inertial reference frame comoving to a given point also does not
  change, otherwise there would be a way of choosing an absolute reference
  frame using a clock.

  Thus at a displacement from an arbitrary reference frame to another one,
  distinct from the first one by its constant position in the coordinate
  system of the first frame, the metrics of space and physical time in the
  new reference frame in the new Cartesian coordinate system do not change
  (are form-invariant). What will change are only the proper characteristics
  of the reference frame, whereas the numerical values and the mathematical
  form of the elements of distance and physical time should remain
  invariable. The standard definitions of the spatial metric and physical
  time
       \begin{equation} \label{ty}
\gamma_{\alpha\beta} =-g_{\alpha \beta } +\frac{g_{0\alpha } g_{0\beta } }{g_{00}}
\end{equation}
     \begin{equation} \label{yu}
\delta\tau =\sqrt{g_{00}}dx^0 +\frac{g_{0\alpha }}{\sqrt{g_{00}}}dx^{\alpha}
\end{equation}
for an arbitrary stationary reference frame give general expressions for
  $dl^2$  and $\delta\tau$ having the form
    \begin{equation} \label{op}
dl^{2} =d\mathbf{R}^{2} +\frac{\left[(\mathbf{\Omega \times R})d\mathbf{R}\right]^{\, 2} }{(1+\mathbf{WR})^{2} -(\mathbf{\Omega \times R})^{2} }=form-invariant
\end{equation}
$$\delta \tau =\sqrt{(1+\mathbf{WR})^{2}-(\mathbf{\Omega \times R})^{2} }dT-$$
\begin{equation}  \label{io}
-\frac{(\mathbf{\Omega \times R})d\mathbf{R}}{\sqrt{(1+\mathbf{WR})^2-(\mathbf{\Omega \times R})^2}}=form-invariant
\end{equation}
An example of using the general form-invariance principle in the form
  (\ref{op}),(\ref{io}) is the full displacement transformation from a uniformly
  rotating reference frame (\ref{W})-(\ref{X}) presented above, and in the form (\ref{op})
  it gives a shortened transformation (i.e., only its spatial part
  (\ref{W})-(\ref{Y})). The proper characteristics of the new reference frame after
  the displacement can be obtained not only from the general form-invariance
  principle in the form (\ref{G}) but also from invariance of the forms (\ref{op}) and
  (\ref{io}). For example, in a rotating reference frame ($\mathbf{W}=0$) the
  length element is
\begin{equation} \label{ui}
dl^{2} =d\mathbf{R}^{2} +\frac{\left[(\mathbf{\Omega \times R})d\mathbf{R}\right]^{\, 2} }{1-(\mathbf{\Omega \times R})^{2} }
\end{equation}
 or 
\begin{equation}  \label{pa}
dl^{2} =dX^{2} +dY^{2} +dZ^{2} +\frac{\Omega ^{2} (XdY-YdX)^{2} }{1-\Omega ^{2} (X^{2} +Y^{2} )}
\end{equation}
At the transformation (\ref{W})-(\ref{Y}), the following equalities hold:
\begin{equation}  \label{as}
 X^{2} +Y^{2} =(b+x)^{2} +\frac{y^{2} }{1-V^{2} }  
\end{equation}
\begin{equation}  \label{sd}
 XdY-YdX=\frac{\left[1-V(V+\Omega x)\right]\, (b+x)dy}{\sqrt{1-V^{2} } } -\frac{\Omega Vy^{2} dy}{\sqrt{1-V^{2} } ^{3} } -\frac{ydx}{\sqrt{1-V^{2} } }  
 \end{equation}
 $$dX^{2} +dY^{2} +dZ^{2}=dx^{2} + \frac{\left[1-V(V+\Omega x)\right]^2dy^2}{1-V^2} + \frac{\Omega^2 V^2y^2 dy^2}{(1-V^2)^2}+$$
 \begin{equation}  \label{dfy}
 +\frac{2\Omega Vydxdy}{1-V^2}+dz^{2}
\end{equation}
Substituting them into (\ref{pa}), we obtain
$$dl^{2}=\left\{\frac{\Omega^2 V^2x^2 dy^2}{1-V^2} -2V\Omega xdy^2-V^2dy^2 +\frac{\Omega ^2 V^2y^2 dy^2}{(1-V^2)^2} +\frac{2\Omega Vydxdy}{1-V^{2} } \right\}+$$
\begin{equation} \label{df}
+dx^{2} +dy^{2} +dz^{2} +\frac{\Omega ^{2} \left(\left[1-V(V+\Omega x)\right]\, (b+x)dy-\frac{\Omega Vy^{2} dy}{1-V^{2} } -ydx\right)^{2} }{(1-V^{2} )\left[1-(V+\Omega x)^{2} \right]^{} -\Omega ^{2} y^{2} }
\end{equation}
Inserting the term in the curly brackets in (\ref{df}) to the numerator of the
  fraction, opening the brackets and collecting the terms, after rather
  bulky calculations we reduce the expression (\ref{df}) to the form
 \begin{equation} \label{fg}
dl^{2} =dx^{2} +dy^{2} +dz^{2} +\frac{\Omega ^{2} (xdy-ydx)^{\, 2} }{\left(1-V(V+\Omega x)\right)^{2} -\Omega ^{2} (x^{2} +y^{2} )}
\end{equation}
 The expression (\ref{fg}), written in a vector form, is equivalent to (\ref{op}),
  where the quantities $\mathbf{w}$ and $\boldsymbol{\omega}$ are given by
  (\ref{we}) and (\ref{Tr}), respectively.

\section{Conclusions}

The role of the general form-invariance principle in SR is that it is a
  kind of ``selection rule'' allowing for singling out from the whole set
  of transformations the true coordinate and time transformations
  corresponding to a transition from one reference frame to another.
  In cases where the coordinate transformations preserve the general forms
  like (\ref{B})-(\ref{D}),(\ref{op}),(\ref{io}), one can assert that such transformations
  between reference frames are not a trivial mathematical substitutions
  of spatial coordinates and time within the same reference frame but are
  connected with physical motion or relative positions. An important
  practical consequence of this principle are the Lorentz--Nelson motion
  transformation (\ref{S}),(\ref{T}), and the displacement transformation (\ref{W})-(\ref{X})
  from a rotating reference frame.

  The hierarchy of principles of SR by their degree of generality is at
  present viewed as follows. A foundation of SR is the principle that
  the geometry is the Minkowski pseudo-Euclidean one. Its direct
  consequence is the general form-invariance principle, and then follow the
  generalized and special relativity principles.

\end {document}